\documentclass{article}%
\usepackage{amsmath}
\usepackage{amsfonts}
\usepackage{amssymb}
\usepackage{graphicx}%
\newcommand{\bpartial}{\mathop{\partial\kern -4pt\raisebox{.8pt}{$|$}}}
\newcommand{\bra}{\mathopen{[\kern-1.6pt[}}
\newcommand{\ket}{\mathclose{]\kern-1.5pt]}}
\newcommand{\bbra}{\mathopen{[\kern-2.2pt[\kern-2.3pt[}}
\newcommand{\bket}{\mathclose{]\kern-2.1pt]\kern-2.3pt]}}

\makeindex
\begin{document}

\title {\large{ \bf $2+1$ dimensional gravity from Maxwell and semi-simple extension of the Poincar\'{e} gauge symmetric models }}

\vspace{3mm}

\author {  \small{ \bf S. Hoseinzadeh }\hspace{-2mm}{ \footnote{ e-mail: hoseinzadeh@azaruniv.edu}} { \small
and}
\small{ \bf A. Rezaei-Aghdam }\hspace{-2mm}{ \footnote{Corresponding author. e-mail:
rezaei-a@azaruniv.edu}} \\
{\small{\em Department of Physics, Faculty of Science, Azarbaijan Shahid Madani University, }}\\
{\small{\em  53714-161, Tabriz, Iran  }}}

\maketitle

\begin{abstract}
We obtain $2+1$ dimensional gravity with cosmological constant
which is coupled to gauge fields, using Maxwell and semi-simple
extension of the Poincar\'{e} gauge symmetric models (i.e.
Chern-Simons models with these gauge groups). Also, we obtain some
$Ads$ and BTZ type solutions for the classical equations of
motion for these $2+1$ dimensional gravities. For the semi-simple
extension of the Poincar\'{e} gauge group we investigate the
$Ads/CFT$ correspondence and show that the model at the boundary
is equivalent to the sum of three WZW models over group
$SO(2,1)$. Then, we show that the central charge of the $CFT$ is
the same as that of $CFT$ at the boundary of $Ads$ spacetime
related to the Chern-Simons model with gauge group $SO(2,2)$. Finally, we show that these two $2+1$ dimensional
gravity models are dual (canonically transformed) to each other.
\end{abstract}
\newpage
\section {\large {\bf Introduction}}
\setlength{\parindent}{0cm}

Maxwell algebra ($\cal{M}$) was introduced four decade ago
\cite{H. Bacry}, \cite{R. Schrader} by replacing commuting
four-momenta in Poincar\'{e} algebra (of 4 dimensional spacetime)
with noncommuting ones; resulting in new Abelian generators.
Nearly at that time the noncommutative four-momenta with the
Lorentz generators\footnote{$[P_{a},P_{b}]\sim J_{ab}$ where
$J_{ab}$ are the Lorentz generators.} (the de Sitter spacetime
algebra) have been applied for unifying a geometric formulation
of gravity and supergravity resulting in the cosmological term
\cite{S. W. MacDowell}. Recently, a generalized cosmological term
was resulted by gauging the Maxwell algebra (without gauge
invariance) \cite{1 J.A. de Azcarraga}. Also, in \cite{D. V.
Soroka} a gauge invariant model by gauging the semi-simple
extension of the Poincar\'{e} algebra\footnote{Here the new
generators resulted from noncommuting four-momentum are non-Abelian
(see (35)).} ($\cal{S}$) has been presented. There are other
applications of Maxwell symmetries of 4 dimensional spacetime
such as its supersymmetrization \cite{S. Bonanos} and
cosmological applications (see for example \cite{2 J.A. de
Azcarraga}). Up to our knowledge, the Maxwell algebra and its
nonabelian extension i.e. the semi-simple extension of the
Poincar\'{e} algebra in $2+1$ dimensional spacetime have not yet been
studied. Here, we will try to study them and obtain $2+1$
dimensional gravity from Maxwell and semi-simple extension of the
Poincar\'{e} gauge symmetric models. Then, we will show that
these models are equivalent to Chern-Simons models over those
gauge groups namely they are exactly soluble models. We will also
obtain some solutions, (black holes and $Ads$) for their
equations of motion; and study the $Ads/CFT$ correspondence for
the last model at the boundary and obtain the central charge of
the $CFT$. The outlines of the paper are as follows:

In section two, after presenting the Maxwell algebra of $2+1$
dimensional spacetime we will try to gauge this symmetries and
obtain the gauge invariant gravitational model as in \cite{1 E.
Witten} where Witten obtained a $2+1$ dimensional gravity by
$ISO(2,1)$ gauge group. Here, the result is a $2+1$ dimensional
gravity without cosmological term which is coupled to Abelian gauge
fields, and similar to \cite{1 E. Witten} this model is
equivalent to Chern-Simons model with Maxwell gauge group i.e. it
is exactly soluble model. Then, we solve the equations of motion
for this model. We obtain flat and BTZ \cite{Banados} type
solutions, such that here we have Abelian gauge fields coupled to
$2+1$ dimensional gravity. In section three, we will try to perform
these works for the semi-simple extension of the Poincar\'{e}
algebra. In section four, we will study the Ads/CFT
correspondence (as \cite{J.D. Brown} and \cite{O. coussaert}) for
the Chern-Simons model with the semi-simple extension of the
Poincar\'{e} gauge group \textbf{($S$)}. We shall show that the
$\cal{S}$ algebra can be rewritten as a direct sum of three
$SO(2,1)$ algebras, and show that at the boundary the C-S action
can be written as a sum of three chiral WZW  models over the group
$SO(2,1)$. Then, we obtain the central charge of the $CFT$ at the
boundary and show that the central charge of the $CFT$ is the
same as that of $CFT$ at the boundary of $Ads$ spacetime related
to the Chern-Simons model with gauge group $SO(2,2)$. Then, we
show that these two $2+1$ dimensional gravities are dual to each other (i.e. we show that they are
canonically transformed to each other). Some concluding remarks
are given in section five.

\section {\large {\bf 2+1 dimensional gravity from Maxwell gauge algebra and Chern-Simons action}}
\setlength{\parindent}{0cm}

~~~~~In this section, we will construct gauge invariant action with Maxwell gauge
group in $2+1$ dimensional spacetime and investigate its relation to $2+1$ dimensional pure gravity
and Chern-Simons action, similarly as Witten obtained a $2+1$ dimensional
gravity from $ISO(2,1)$ gauge group in \cite{1 E. Witten}. Let us
first consider the commutation relations for the Poincar\'{e} algebra in
$2+1$ dimensional spacetime
\begin{eqnarray} \label{L1}
[J_{a},J_{b}] = \epsilon_{abc}J^{c},~~~~~~~~  [J_{a},P_{b}] = \epsilon_{abc}P^{c},~~~~~~~~  [P_{a},P_{b}] = 0,
\end{eqnarray}
where  $J_{a}$  and  $P_{a}$ ($a=0,1,2$) are generators of
rotation and translation in spacetime.\footnote{The rotation
generators  have the form $J_{ab}$, here we use the
$J^{a}=\frac{1}{2}\epsilon^{abc}J_{bc}$ form for these
generators.} As for the $3+1$ dimensional spacetime \cite{H.
Bacry} (see also \cite{1 J.A. de Azcarraga}) one can write the
$D=2+1$ nine dimensional Maxwell algebra $ {\cal
M}=(J_{a},P_{a},Z_{a}) $ by a noncommutative modification of the
Abelian three-momenta commutators in the Poincar\'{e} algebra as
follows:\footnote{Note that the commutator $[J_{a},Z_{b}]$ can be
obtained from Jacobi identity.}
\begin{eqnarray}   \nonumber
~~~~~~~[J_{a},J_{b}] = \epsilon_{abc}J^{c},~~~~~~~~
[J_{a},P_{b}] = \epsilon_{abc}P^{c},~~~~~~~~  [P_{a},P_{b}] =
\Lambda \epsilon_{abc}Z^{c} ,
\end{eqnarray}
\begin{eqnarray}\label{L2}
~~~~~~[J_{a},Z_{b}] = \epsilon_{abc}Z^{c},~~~~~~~  [P_{a},Z_{b}] = 0,~~~~~~~~~~~~~~~  [Z_{a},Z_{b}] = 0,~~~~~~~
\end{eqnarray}
where $Z_{a}$'s are new generators and $ \Lambda $ is a constant. We see that
${\cal I}=\{Z_{a}\}$ is an Abelian ideal of the Maxwell algebra $\cal{M}$,
hence the $D=2+1$ nine dimensional Maxwell Lie algebra is nonsemi-simple and indeed
it is an algebraic Lie algebra i.e., $[{\cal M},{\cal M}]={\cal M}$.
Now, to obtain a gauge invariant action with $D=2+1$ Maxwell gauge group we need to
construct a gauge field which is Maxwell algebra valued one form as follows:
\begin{eqnarray} \nonumber
h=h_{i}~dx^{i},
\end{eqnarray}
\begin{eqnarray}\label{L3}
h_{i}=h_{i}^{~B}X_{B}=e_{i}^{~a}P_{a}+\omega_{i}^{~a}J_{a}+A_{i}^{~a}Z_{a},
\end{eqnarray}
where $i,j=0,1,2$ are spacetime indices such that the one form fields are defined as follows:
\begin{eqnarray}\label{L4}
e^{a}=e_{i}^{~a} dx^{i} ~~,~~ \omega^{a}=\omega_{i}^{~a} dx^{i} ~~,~~ A^{a}=A_{i}^{~a} dx^{i},
\end{eqnarray}
where $ e_{i}^{~a} , \omega_{i}^{~a} $ are vierbein and spin
connection, respectively. Furthermore, here we have new Abelian
gauge fields $ A_{i}^{~a} $. To obtain the gauge transformations
of these gauge fields we use the following infinitesimal gauge
parameter:
\begin{eqnarray}\label{L5}
u=\rho^{a}P_{a}+\tau^{a}J_{a}+\lambda^{a}Z_{a}.
\end{eqnarray}
In this way, using the following relation for the gauge transformations:
\begin{eqnarray}\label{L6}
h_{i} \rightarrow h^{\prime}_{i}=U^{-1}h_{i}U+U^{-1}\partial_{i}U ,
\end{eqnarray}
with  ~ $  U = e^{-u} \; \simeq  \; 1-u  $   ~ and ~   $  U^{-1} = e^{u} \; \simeq  \; 1+u , $
~we obtain the following transformations for the gauge fields:
\begin{eqnarray}    \nonumber
~~~~~~~~~~~~~~~~\delta e_{i}^{\
a}=-\partial_{i}\rho^{a}-\epsilon^{abc} e_{ib} \;
\tau_{c}-\epsilon^{abc}\omega_{ib} \; \rho_{c} ~,
\end{eqnarray}
\begin{eqnarray}\label{L7}
\delta \omega_{i}^{\ a}=-\partial_{i}\tau^{a}-\epsilon^{abc}\omega_{ib} \; \tau_{c} ~,
\end{eqnarray}
\begin{eqnarray}     \nonumber
~~~~~~~~~~~~~~~~~~~~~~~~~~~~~~~~~\delta A_{i}^{\
a}=-\partial_{i}\lambda^{a}-\Lambda  \epsilon^{abc} e_{ib} \;
\rho_{c}-\epsilon^{abc}\omega_{ib} \; \lambda_{c}
-\epsilon^{abc}A_{ib} \; \tau_{c}  ~.
\end{eqnarray}
As we expect, the gauge transformations of the vierbein and spin
connection are the same as that of \cite{1 E. Witten}, and only
here we have new gauge fields and their transformations. Now, to
write the topological and gauge invariant action of the form
$\int {\cal R}^{A} \wedge {\cal R}^{B} \Omega_{AB}$ \cite{1 E.
Witten}, where $\Omega_{AB}$ is an ad-invariant metric on the
Maxwell algebra $\cal{M}$, we need to calculate Ricci curvature
as follows:
\begin{eqnarray}\label{L8}
\mathcal{R}=\mathcal{R}_{ij} dx^{i} \wedge dx^{j} =
\mathcal{R}^{A} X_{A} = \mathcal{R}_{ij}^{~A} X_{A} dx^{i} \wedge
dx^{j} ,
\end{eqnarray}
\begin{eqnarray}\label{L9}
\mathcal{R}_{ij}=\partial_{[i}h_{j]}+[h_{i},h_{j}] =
\mathcal{R}_{ij}^{~A} X_{A}
=T_{ij}^{~~a}~P_{a}+R_{ij}^{~~a}~J_{a}+F_{ij}^{~~a}~Z_{a} ~,
\end{eqnarray}
such that for the torsion $ T_{ij}^{~~a} $, standard Riemannian curvature $ R_{ij}^{~~a} $ and the
new field strength $ F_{ij}^{~~a} $ we have:
\begin{eqnarray}  \nonumber
~~~~~~~~~~~~~T_{ij}^{~~c}=\partial_{[i}
e_{j]}^{~c}+\epsilon_{ab}^{~~c}(e_{i}^{\ a} \omega_{j}^{\
b}+\omega_{i}^{\ a} e_{j}^{\ b}),
\end{eqnarray}
\begin{eqnarray}\label{L10}
R_{ij}^{~~c}=\partial_{[i} \omega_{j]}^{~c}+\epsilon_{ab}^{~~c}
\omega_{i}^{\ a} \omega_{j}^{\ b},
\end{eqnarray}
\begin{eqnarray}  \nonumber
~~~~~~~~~~~~~~~~~~~~~~~~~~~~~F_{ij}^{~~c}=\partial_{[i}
A_{j]}^{~c} + \epsilon_{ab}^{~~c}(\Lambda e_{i}^{\ a} e_{j}^{\ b}
+ \omega_{i}^{\ a} A_{j}^{\ b}+A_{i}^{\ a} \omega_{j}^{\ b}).
\end{eqnarray}
Furthermore, using the relation ~$ f_{AB}^{~~~C}
~\Omega_{CD}+f_{AD}^{~~~C}~\Omega_{CB}=0 $~ \footnote{$ f_{AB}^{~~~C} $ is the
structure constant of the Maxwell Lie algebra $\cal{M}$ \eqref{L2}.} \cite{C. Nappi}
the ad-invariant metric $ \Omega_{AB}=\langle X_{A},X_{B} \rangle$ of the
Maxwell algebra can be obtained as follows:
\begin{eqnarray}
\langle J_{a},J_{b} \rangle ~=~\alpha\: \eta_{ab},  \hspace{.5cm}
\langle J_{a},P_{b} \rangle ~=~\beta\: \eta_{ab},  \hspace{.5cm}
\langle J_{a},Z_{b} \rangle ~=~\gamma\: \eta_{ab},            \nonumber
\end{eqnarray}
\begin{eqnarray}\label{L11}
\langle P_{a},P_{b} \rangle ~=~\Lambda~\gamma\: \eta_{ab},  \hspace{.5cm}
\langle P_{a},Z_{b} \rangle ~=~ \langle Z_{a},Z_{b} \rangle ~=~0,
\end{eqnarray}
where $ \eta_{ab} $ is the three dimensional Minkowski metric and $\alpha,\beta$ and $\gamma$ are real constant
parameters (with $\gamma \neq0$ such that $det\Omega_{AB}=-\Lambda^3 \gamma^9$).
Note that for $\Lambda=0$ this metric is degenerate; hence for the Poincar\'{e} algebra one must use
the standard ad-invariant metric.
Using this metric, one can construct the following quadratic Casimir operator
\begin{eqnarray}\nonumber
W=X_{A}~\Omega^{AB}X_{B}=\frac{2}{\gamma}J^{a}Z_{a}+\frac{1}{\Lambda
\gamma}P^{a}P_{a}-\frac{2\beta}{\Lambda
\gamma^{2}}P^{a}Z_{a}+\frac{(\beta^2-\alpha\Lambda\gamma)}{\Lambda
\gamma^{3}}Z^{a}Z_{a},
\end{eqnarray}
where $\Omega^{AB}$ is inverse of the ad-invariant metric.
Now, in this way one can construct the topological and gauge invariant action
in the following form:

\begin{eqnarray}\label{L12}
\hspace{-6.5cm}            I=\frac{1}{16\pi} \int_{Y} \mathcal{R}^{A}\wedge
\mathcal{R}^{B} ~\Omega_{AB}=\frac{1}{16\pi} \int_{Y} d^{4}x ~ \epsilon^{ijkl}
\langle \mathcal{R}_{ij}~,~\mathcal{R}_{kl} \rangle
\end{eqnarray}
\vspace{-4mm}
\begin{eqnarray}\label{L13}
=\frac{1}{16\pi} \int_{Y} d^{4}x ~
\epsilon^{ijkl} ~ (\Lambda \gamma~ T_{ij}^{~~ c}~ T_{klc}+
\alpha~ R_{ij}^{~~ c}~ R_{klc}  + 2\gamma~ R_{ij}^{~~ c}~
F_{klc}  + 2\beta~ T_{ij}^{~~ c}{} R_{klc} ),~~~~~~~~~~~~~~~~~~~~~~~~
\end{eqnarray}
where Y is a four dimensional manifold with boundary $ M=\partial
Y$. Now using \eqref{L10} and integration by part one can rewrite this action as
the following one:
\begin{eqnarray}\label{L14}
I= \frac{1}{8\pi} \int_{M} d^{3}x ~  \epsilon^{ijk}~ \Big[ 2\beta ~ e_{ic}~
D_{j} \omega_{k}^{~c} + \alpha~
\omega_{ic}~(\partial_{j}~ \omega_{k}^{\ c}-\partial_{k}~
\omega_{j}^{\ c} + \frac{2}{3}~ \epsilon^{abc}~ \omega_{ja}~
\omega_{kb})~~~~~~~~~~~~~~~~~~~~~~~~~~~~~~     \nonumber     \\
+\Lambda \gamma~e_{ic}~D_{j} e_{k}^{~c}
+ 2\gamma~ \omega_{ic}~(\partial_{j}~ A_{k}^{\ c}-\partial_{k}~
A_{j}^{\ c} + \epsilon^{abc}~ \omega_{ja}~ A_{kb})  \Big],~~~~~~~~~~~~~~~~~~~~~~~~~~~~
\end{eqnarray}
where
\begin{eqnarray} \label{L15}
D_{j} e_{k}^{\ a}= \partial_{[j}
e_{k]}^{~a}+\epsilon_{bc}^{~~a}(e_{j}^{\ b} \omega_{k}^{\
c}+\omega_{j}^{\ b} e_{k}^{\ c}),
\end{eqnarray}
\begin{eqnarray} \label{L16}
~D_{j} \omega_{k}^{\ a}= \partial_{[j}
\omega_{k]}^{~a}+\epsilon_{bc}^{~~a}\omega_{j}^{\ b} \omega_{k}^{\
c}.~~~~~~~~~~~~~~
\end{eqnarray}
Note that this action is the Chern-Simons action; i.e. by use of
the following Chern-Simons action:
\begin{eqnarray}\label{L17}
I_{cs}=\frac{1}{4\pi}\int_{M} \Big(\langle h \wedge dh \rangle +
\frac{1}{3}~\langle h \wedge [h \wedge h]\rangle \Big),
\end{eqnarray}
and using \eqref{L3} and \eqref{L11} one can obtain \eqref{L14}; in this way the action
\eqref{L14} is an exactly soluble model. We see
that the first term of the action \eqref{L14} is the pure gravity
(Einstein-Hilbert action) \cite{1 E. Witten} and the second term together with the third one
(with $\gamma=\alpha$) are the Chern-Simons action for the gauge group $SO(2,2)$ or
$SO(3,1)$ \cite{1 E. Witten}. The fourth term is a new one which
represents the coupling of spin connection to the gauge fields $A_{i}^{\ a}$. Note that
there is no kinetic term for the new Abelian gauge fields $A_{i}^{\ a}$; this is because
the $\langle Z_{a},Z_{b} \rangle$ element of the ad-invariant metric is zero. Hence, if one adds the
kinetic term of the gauge fields $A_{i}^{\ a}$ to the action \eqref{L14}, then it is not
a gauge invariant model.

Now, in the following, we consider the model \eqref{L14} or
\eqref{L17} as a gauge invariant model (invariant under
transformations \eqref{L7}) over three dimensional spacetime
(with boundary) $M$ and try to obtain the equations of motion and
solve them. The equations of motion for the action \eqref{L14}
can be obtained as follows;

the equations of motion for the fields  $ e_{ia} $ have the following form:
\begin{eqnarray}\label{L18}
\epsilon^{ijk}{}(\Lambda
\gamma~ D_{j} e_{k}^{\ a} + \beta ~ D_{j} \omega_{k}^{\ a} )=0 ,
\end{eqnarray}
the equations of motion with respect to  $ \omega_{ia} $ are as follows:
\begin{eqnarray}\label{L19}
\epsilon^{ijk}{}[\alpha ~ D_{j} \omega_{k}^{\ a} + \gamma ~ (D_{j} A_{k}^{\
a}+ \Lambda{} \epsilon^{abc} e_{jb} e_{kc}) + \beta ~ D_{j} e_{k}^{\ a}
]=0,
\end{eqnarray}
where
\begin{eqnarray} \label{L20}
D_{j} A_{k}^{\ a}= \partial_{[j} A_{k]}^{~a}+\epsilon_{bc}^{~~a}
(A_{j}^{\ b} \omega_{k}^{\ c}+\omega_{j}^{\ b} A_{k}^{\ c}),
\end{eqnarray}
and finally the equations of motion with respect to  $  A_{ia} $ have the following form:
\begin{eqnarray} \label{L21}
\epsilon^{ijk}{} D_{j} \omega_{k}^{\ a}=0,
\end{eqnarray}
such that using \eqref{L10} one can rewrite the above equations as follows:
\begin{eqnarray}
\epsilon^{ijk} ~ T_{jk}^{\quad a}=0,     \label{L22}   \\
\epsilon^{ijk} ~ R_{jk}^{\quad a}=0,     \label{L23}   \\
\epsilon^{ijk} ~ F_{jk}^{\quad a}=0.     \label{L24}
\end{eqnarray}
We see that like $SO(2,2)$ and $SO(3,1)$ Chern-Simons actions in
\cite{1 E. Witten}, the equations of motion of the action \eqref{L14}
can be rewritten as a zero's of the field strengths.
Now, in the following we will try to obtain different solutions for these equations.
\subsection  {Solutions of the equations of motion for the
Chern-Simons action with Maxwell gauge group}
Here, we apply two ansatzes to obtain the solutions of the
equations \eqref{L22}-\eqref{L24}; i.e. flat and BTZ type solutions.
\subsubsection  {Flat solution}
We use the following ansatz for the metric in the equations \eqref{L22}-\eqref{L24}:
\begin{eqnarray}\label{L25}
ds^2=-N^2(r) dt^2+ \frac{1}{N^2(r)} dr^2 +r^2  d \phi^{2},
\end{eqnarray}
where $\{x^{0},x^{1},x^{2}\}=\{t,r,\phi\}$ are the coordinates of the spacetime.
After some calculations one can obtain
\begin{equation}          \nonumber
N(r) = C_{3}, ~~~~~~~~~~~~~~~~~~~~~~~~~~~~~~~~~~~~~~~~~~~~~~~~~~~~~~~~~
\end{equation}
\begin{equation}          \nonumber
\omega^{0}(r)= - C_{3}~d\varphi, ~~~~ \omega^{1}(r)=0, ~~~~~~~ \omega^{2}(r)=0,~~~~~~~~~~~
\end{equation}
\begin{equation} \label{L26}
A^{0}(r)=C_{2}~dt + f(r)~dr + \Big(-\frac{\Lambda r^{2}}{2 C_{3}} + C_{1} \Big)~d\varphi,~~~~~~~~~~~~
\end{equation}
\begin{equation}  \nonumber
A^{1}(r)=\frac{1}{C_{3}} g^{\prime}(r)~dr + h(r)~d\varphi, ~~~~~~~~~~~~~~~~~~~~~~~~~~~~~~~~~
\end{equation}
\begin{equation}   \nonumber
A^{2}(r)=- \Lambda r~dt - \frac{1}{C_{3}} h^{\prime}(r)~dr + g(r)~d\varphi,~~~~~~~~~~~~~~~~~~~~~
\end{equation}
where $C_{1}$, $C_{2}$ and $C_{3}$ are real constants and $f(r)$,
$g(r)$ and $h(r)$ are arbitrary functions of ~$r$ and prime shows
the derivative with respect to $r$.

\subsubsection  {BTZ-type solution}
Here, we use the following BTZ-type ansatz \cite{Banados} for the
metric in the equations \eqref{L22}-\eqref{L24}:
\begin{eqnarray}\label{L27}
ds^2=-N^2(r) dt^2+ \frac{1}{N^2(r)} dr^2 +r^2 (N^{\phi}(r) ~ dt + d \phi)^2,
\end{eqnarray}
after some calculations one can obtain the following solution for these equations:
\begin{eqnarray}            \nonumber
N^2(r) = -\frac{D_{3}}{r^2}-\frac{M}{2},~~~~~~~~~~~~~~~~~
N^{\phi}(r) = \frac{\sqrt{-D_{3}}}{r^2},~~~~~~~~~~~~~~~~~~~
\nonumber
\end{eqnarray}
\begin{equation}          \nonumber
\omega^{0}(r) = N(r) ~d\varphi, ~~~~~~~
\omega^{1}(r) = r N^{\phi}(r) ~d\varphi, ~~~~~~~
\omega^{2}(r) = -\frac{N^{\phi}(r)}{N(r)}~dr,
\end{equation}
\begin{equation} \label{L28}
A^{0} = \frac{D_{2}}{\sqrt{-D_{3}}} N(r) ~dt + h(r)~dr
+ \frac{q(r)}{N(r)} ~d\varphi,~~~~~~~~~~~~~~~~~~~~~~~~~~~~~~~~~~~~
\end{equation}
\begin{equation}          \nonumber
A^{1} = \Big( \Lambda r + \frac{D_{2}}{r} \Big)~dt + \Big( \frac{rN^{\phi}(r) h(r)
- g^{\prime}(r)}{N(r)} \Big) ~dr + f(r)~d\varphi, ~~~~~~~~~~~~~~~
\end{equation}
\begin{equation}          \nonumber
A^{2} = \Big( \frac{f^{\prime}(r)}{N(r)}
+  \frac{N^{\phi}(r)}{N^{3}(r)}~q(r) \Big)~dr + g(r)~d\varphi, ~~~~~~~~~~~~~~~~~~~~~~~~~~~~~~~~~~~~
\end{equation}
where
\begin{equation}          \nonumber
q(r)= \frac{\Lambda}{2}r^2+rN^{\phi}(r)f(r) + D_{1},
\end{equation}
$D_{1}$, $D_{2}$, $D_{3}$ and $M$ are real constants and $f(r)$,
$g(r)$ and $h(r)$ are arbitrary functions of ~$r$.
In this way, we have obtained the BTZ type solution \cite{Banados} such that
it is coupled to the gauge field matter (\eqref{L14} and \eqref{L28}).  \\
To determine the constants of the solutions, we use the
energy-momentum tensor at the boundary. The quasilocal stress
tensor defined locally on the boundary of a given spacetime
region is as follows: \cite{J.D. Brown and J.W. York}, \cite{V.
Balasubramanian}
\begin{equation} \label{L29}
T^{ij}= \frac{2}{\sqrt{-\gamma}} \frac{\delta I}{\delta
\gamma_{ij}}= \frac{2}{\sqrt{-\gamma}} \frac{\delta I}{\delta
e_{\ell}^{~d}} \frac{\delta e_{\ell}^{~d}}{\delta \gamma_{ij}},
\end{equation}
where $\gamma_{ij}$ is the boundary metric. The boundary
$\partial M_{r}$ of our spacetime $M$ is a cylindrical shell at
fixed $r$. Varying the action produces a bulk term which is zero
using the equations of motion, plus a boundary term as:
\begin{equation} \label{L30}
\delta I= -\frac{1}{4\pi}~ \delta \int_{\partial M_{r}} d^{2}x~
\epsilon^{ij} \Big[\alpha~ \omega_{ic}\omega_{j}^{~c} + 2\beta~
e_{ic}\omega_{j}^{~c} + 2\gamma~ \omega_{ic}A_{j}^{~c} + \Lambda
\gamma~ e_{ic}e_{j}^{~c} \Big] + \int_{\partial M_{r}} d^{2}x~
\frac{\delta I_{ct}}{\delta \gamma_{ij}} \delta \gamma_{ij},
\end{equation}
where $\epsilon^{20}=+1$ and $I_{ct}$ is the counterterm action
which is added in order to obtain a finite stress tensor at
$r \rightarrow \infty$ \cite{V. Balasubramanian}. Then, we get the
quasilocal stress tensor for this model as:
\begin{equation} \label{L31}
T^{ij}= - \frac{1}{2\pi} \frac{\beta}{\sqrt{-\gamma}}
\epsilon^{in} \omega_{n}^{~c} \gamma^{jk} e_{kc},
\end{equation}
where $\sqrt{-\gamma}=r N(r)$ and $\gamma^{jk}$ is the inverse
boundary metric. For this model, we have $I_{ct}=0$ such
that using the above solution, the components of quasilocal stress
tensor are obtained as follows:
\begin{equation} \label{L32}
T^{00}= \frac{\beta}{2\pi rN(r)},~~~~~~~~~~~ T^{02}=T^{20}=T^{22}=0.
\end{equation}
The mass and angular momentum which are the conserved charges
associated with time translation and rotation respectively, have
been defined in \cite{V. Balasubramanian} as:
\begin{equation}         \nonumber
m= \int_{0}^{2\pi} d\varphi~ r N(r) u^{i}u^{j}T_{ij},
\end{equation}
\begin{equation}  \label{L33}
P_{\varphi}=  \int_{0}^{2\pi} d\varphi~ r^{3} \gamma_{ij} u^{i} T^{2j},~~~~
\end{equation}
where $u^{i}=\frac{1}{\sqrt{-N^{2}(r)+r^{2}(N^{\phi}(r))^{2}}}
~\delta^{i,0}$ is the timelike unit normal to spacelike surface
$\Sigma$ in $\partial \mathcal{M}$. After some calculations we
find that the mass and angular momentum have the following forms:
\begin{equation} \label{L34}
m=\frac{\beta}{2} M ,~~~~~~~ P_{\varphi}=0.
\end{equation}

The above metric has a singularity at $r =
\sqrt{\frac{-2D_{3}}{M}}$.  This singularity is not a curvature
singularity, but a coordinate one associated with horizon in the
Schwarzschild-type spacetime, and as is well known, there is
other coordinate system for which this type of singularity is
removed. It describes a non-rotating $(P_{\varphi}=0)$ black hole
in $2+1$ dimensional spacetime with mass $M$.

\section {\large {\bf 2+1 dimensional gravity from semi-simple extension
of the Poincar\'{e} gauge Algebra and Chern-Simons action}}
Here, we try to perform calculations similar to the section two;
for the semi-simple extension of the Poincar\'{e} algebra. This
algebra in $D=2+1$ dimensional spacetime can be obtained from
Maxwell's one by deforming the commutator of the generator
$Z_{a}$ in \eqref{L2} as follows:
\begin{eqnarray}   \nonumber
[J_{a},J_{b}] = \epsilon_{abc} J^{c}, ~~~~~~[J_{a},P_{b}] =
\epsilon_{abc} P^{c}, ~~~~~~[P_{a},P_{b}] = k \epsilon_{abc}
Z^{c},
\end{eqnarray}
\begin{eqnarray} \label{L35}
[J_{a},Z_{b}] = \epsilon_{abc} Z^{c}, ~~~~~~[P_{a},Z_{b}] =
-\frac{\Lambda}{k} \epsilon_{abc} P^{c}, ~~~~~~[Z_{a},Z_{b}] =
-\frac{\Lambda}{k} \epsilon_{abc} Z^{c},
\end{eqnarray}
where $k$ is a constant. The commutator of $[P_{a},Z_{b}]$ can be obtained
from Jacobi identities. Note that this algebra is a semi-simple one.
For the Chern-Simons model, the gauge field can be written similarly
as \eqref{L3} and \eqref{L4}. The gauge transformations \eqref{L7} are deformed as follows:
\begin{eqnarray}    \nonumber
\delta e_{i}^{\ a}=-\partial_{i} \rho^{a}-\epsilon^{abc} e_{ib}
~ \tau_{c}-\epsilon^{abc}\omega_{ib} ~ \rho_{c}
+\frac{\Lambda}{k} \epsilon^{abc}e_{ib} ~ \lambda_{c}
+\frac{\Lambda}{k} \epsilon^{abc}A_{ib} ~ \rho_{c},~~
\end{eqnarray}
\begin{eqnarray}\label{L36}
\hspace{-6.4cm}    \delta \omega_{i}^{\ a}=-\partial_{i}
\tau^{a}-\epsilon^{abc}\omega_{ib} ~ \tau_{c},~
\end{eqnarray}
\begin{eqnarray}     \nonumber
~~\delta A_{i}^{\ a}=-\partial_{i} \lambda^{a}- k \epsilon^{abc}
e_{ib}  ~ \rho_{c}-\epsilon^{abc}\omega_{ib} ~ \lambda_{c}
-\epsilon^{abc}A_{ib} ~ \tau_{c} +\frac{\Lambda}{k}
\epsilon^{abc}A_{ib} ~ \lambda_{c}.~~~~~
\end{eqnarray}
Furthermore, one can obtain the field strengths in the same way of section two as follows:
\begin{eqnarray}  \nonumber
T_{ij}^{~~c}=\partial_{[i}
e_{j]}^{~c}+\epsilon_{ab}^{~~c}(e_{i}^{\ a} \omega_{j}^{\ b}
+\omega_{i}^{\ a} e_{j}^{\ b}) -\frac{\Lambda}{k}
\epsilon_{ab}^{~~c} ( e_{i}^{a} ~ A_{j}^{b}+ A_{i}^{a} \;
e_{j}^{b}),~~~
\end{eqnarray}
\begin{eqnarray}\label{L37}
R_{ij}^{~~c}=\partial_{[i} \omega_{j]}^{~c}+\epsilon_{ab}^{~~c}
~  \omega_{i}^{\ a} \omega_{j}^{\
b},~~~~~~~~~~~~~~~~~~~~~~~~~~~~~~~~~~~~~~~~~~~~~~~
\end{eqnarray}
\begin{eqnarray}  \nonumber
F_{ij}^{~~c}=\partial_{[i} A_{j]}^{~c} + \epsilon_{ab}^{~~c} (k ~
e_{i}^{\ a} e_{j}^{\ b} + \omega_{i}^{\ a} A_{j}^{\ b}+A_{i}^{\
a} \omega_{j}^{\ b}) -\frac{\Lambda}{k} \epsilon_{ab}^{~~c}
A_{i}^{a} ~ A_{j}^{b}.
\end{eqnarray}
The ad-invariant metric can also be obtained as follows:
\begin{eqnarray}
\langle J_{a},J_{b} \rangle ~=~a ~ \eta_{ab}            \hspace{.5cm}  ,  \hspace{.5cm}
\langle J_{a},P_{b} \rangle ~=~b ~ \eta_{ab}            \hspace{.5cm}  ,  \hspace{.5cm}
\langle J_{a},Z_{b} \rangle ~=~d ~ \eta_{ab}            \nonumber
\end{eqnarray}
\begin{eqnarray} \label{L38}
\langle P_{a},P_{b} \rangle ~=~k  d ~ \eta_{ab}                 \hspace{.5cm}  ,  \hspace{.5cm}
\langle P_{a},Z_{b} \rangle ~=~-\frac{\Lambda}{k} b ~ \eta_{ab}      \hspace{.5cm}  ,  \hspace{.5cm}
\langle Z_{a},Z_{b} \rangle ~=~-\frac{\Lambda}{k} d ~ \eta_{ab},
\end{eqnarray}
where $a, b$ and $d$ are arbitrary real constants. As we expect, for the
limiting case $\Lambda \rightarrow 0$, this metric reduces to \eqref{L11}
with $a=\alpha$, $b=\beta$, $d=\gamma$.
The quadratic Casimir operator for this algebra is
\begin{eqnarray}\nonumber
W=X_{A}~\Omega^{AB}X_{B}
=\frac{1}{\frac{\Lambda}{k}a+d} \Big(\frac{\Lambda}{k}J^{a}J_{a}+2J^{a}Z_{a} \Big)
+\frac{1}{\frac{\Lambda}{k}b^2+k d^2} \Big(dP^{a}P_{a} -2bP^{a}Z_{a} \Big)
\end{eqnarray}
\begin{eqnarray}\nonumber
+\frac{(b^2-kda)}{(\frac{\Lambda}{k}a+d)(\frac{\Lambda}{k}b^2+kd^2)}Z^{a}Z_{a}.~~~~~~~~~~~~~~~~~~~~~~~~~~
\end{eqnarray}
Now, with this information one can obtain the topological
invariant action in terms of the field strengths, as follows:
\begin{eqnarray}     \nonumber
\hspace{-2.5cm}     I=\frac{1}{16\pi} \int_{Y} {\cal
R}^{A}\wedge  {\cal R}^{B} ~\Omega_{AB}=\frac{1}{16\pi} \int_{Y}
d^{4}x ~ \epsilon^{ijkl} \langle {\cal R}_{ij}~,~{\cal R}_{kl}
\rangle
\end{eqnarray}
\begin{eqnarray} \label{L39}
=\frac{1}{16\pi} \int_{Y} d^{4}x ~ \epsilon^{ijkl} ~ (k \, d ~ T_{ij}^{~~c}~ T_{kl,c} + 2~b ~
T_{ij}^{~~c}~ R_{kl,c} -2~b~\frac{\Lambda}{k}~T_{ij}^{~~c}~ F_{kl,c}   \nonumber   \\
+ a ~ R_{ij}^{~~c}~ R_{kl,c} +2~d ~ R_{ij}^{~~c}~ F_{kl,c} - d ~ \frac{\Lambda}{k} ~ F_{ij}^{~~c}~ F_{kl,c}).
\end{eqnarray}
Then, replacing from \eqref{L37} and after integration by part;
one can obtain the following action:
\begin{eqnarray} \label{L40}
I=\frac{1}{8\pi} \int_{M} d^{3}x ~\epsilon^{ijk} \Big\{ 2b~e_{ic}
(D_{j} \omega_{k}^{~c}  -\frac{1}{3} \Lambda  \epsilon^{abc}
e_{ja} ~ e_{kb})
+a~\omega_{ic} (\partial_{j}~ \omega_{k}^{\ c}-\partial_{k}~ \omega_{j}^{\ c}
+ \frac{2}{3}~ \epsilon^{abc} \omega_{ja} ~ \omega_{kb})     \nonumber     \\
-2b~\frac{\Lambda}{k}~e_{ic} (D_{j} A_{k}^{\ c} -\frac{\Lambda}{k} ~ \epsilon^{abc} A_{ja} ~ A_{kb})
+2d ~ \omega_{ic} (\partial_{j}~ A_{k}^{\ c}-\partial_{k}~ A_{j}^{~c}
+ \epsilon^{abc} \omega_{ja}~ A_{kb}) \nonumber\\
+ k d ~ e_{ic}~D_{j} e_{k}^{~c} -d~\frac{\Lambda}{k}  A_{ic}
(D_{j} A_{k}^{~c} +2k~ \epsilon^{abc} e_{ja}~ e_{kb}
-\frac{2}{3} ~\frac{\Lambda}{k} ~ \epsilon^{abc} A_{ja}~
A_{kb})~  \Big\}.
\end{eqnarray}
Similar to the previous section, this action is the Chern-Simons
action \eqref{L17}  with the semi-simple extension
of the Poincar\'{e} gauge group \textbf{($\mathcal{S}$)}. Here, in
addition to the previous terms in \eqref{L14}, the cosmological
constant term is explicitly appeared in the lagrangian with the
cosmological constant $\Lambda=-\frac{1}{\ell^{2}}$. Furthermore,
there are new terms which represent the interaction of the
non-Abelian gauge fields $A_{k}^{\ a}$ with each other (in the
form of Chern-Simons terms for the $A_{k}^{\ a}$ fields) and spin
connections and vierbeins. This action is invariant under the
gauge transformations \eqref{L36}.
The equations of motion for the action \eqref{L40} can be obtained as follows.   \\
The equations of motion with respect to $e_{ia}$ have the following form:
\begin{eqnarray} \label{L41}
\epsilon^{ijk} \Big[ k d~ (D_{j} e_{k}^{\ a} -2~\frac{\Lambda}{k}
~  \epsilon^{abc}~ e_{jb} ~ A_{kc} ) + b~ \Big(D_{j}
\omega_{k}^{~a} -\frac{\Lambda}{k} (D_{j} A_{k}^{~a} + k ~
\epsilon^{abc} ~ e_{jb} ~ e_{kc} -\frac{\Lambda}{k} ~
\epsilon^{abc}{} A_{jb} ~ A_{kc} ) \Big) \Big]=0,
\end{eqnarray}
furthermore the equations of motion for $\omega_{ia}$ are as follows:
\begin{eqnarray}  \label{L42}
\epsilon^{ijk} \Big[a \, D_{j} \omega_{k}^{~a} + d ~ \Big(D_{j}
A_{k}^{~a}+ k~ \epsilon^{abc} e_{jb}~ e_{kc} -\frac{\Lambda}{k} ~
\epsilon^{abc}~ A_{jb}~ A_{kc} \Big) + b ~ \Big(D_{j} e_{k}^{~a}
-2~ \frac{\Lambda}{k}~ \epsilon^{abc}~ e_{jb} ~ A_{kc} \Big)
\Big]=0,~~~~~~~
\end{eqnarray}
and finally the equations of motion with respect to $A_{ia}$ have the form:
\begin{eqnarray} \label{L43}
\epsilon^{ijk} \Big[-b ~ \frac{\Lambda}{k} ~(D_{j} e_{k}^{~a}
-2~ \frac{\Lambda}{k} ~ \epsilon^{abc}~ e_{jb} ~ e_{kc} ) +d ~
\Big(D_{j} \omega_{k}^{~a} -\frac{\Lambda}{k} ~ (D_{j} A_{k}^{~a}
+k~ \epsilon^{abc} e_{jb} ~ e_{kc} -\frac{\Lambda}{k}~
\epsilon^{abc}~ A_{jb}~ A_{kc} ) \Big) \Big]=0.
\end{eqnarray}
As in the previous section, one can rewrite these equations in terms of field strengths as follows:
\begin{eqnarray}\label{L44}
\epsilon^{ijk} \Big[k d ~ T_{jk}^{~~a} + b \Big( R_{jk}^{~~a}
-\frac{\Lambda}{k} ~ F_{jk}^{~~a} \Big) \Big]=0,~~~~
\end{eqnarray}
\begin{eqnarray}\label{L45}
\epsilon^{ijk} ~ \Big(a ~ R_{jk}^{~~a} + b ~ T_{jk}^{~~a} + d ~ F_{jk}^{~~a} \Big)=0,~~~~~~~~
\end{eqnarray}
\begin{eqnarray}\label{L46}
\epsilon^{ijk} \Big[-b~ \frac{\Lambda}{k}~ T_{jk}^{~~a}  + d
\Big( R_{jk}^{~~a} -\frac{\Lambda}{k}~ F_{jk}^{~~a} \Big) \Big]=0.
\end{eqnarray}
\subsection  {Solutions of the equations of motion}
As for the previous section, we apply two type ansatzes for the
metric solution of equations \eqref{L44} - \eqref{L46}.
\subsubsection  {Ads-type solution}
If we use the ansatz \eqref{L25} for the metric in the equations
\eqref{L44} - \eqref{L46}; then after some calculations we obtain
the following solutions:
\begin{equation}       \nonumber
N^{2}(r) =1-\Lambda r^{2} , ~~~~~~~~~   \omega^{0} = \zeta(r)
\Big(C_{2}dt + d\varphi  + \frac{f(r)}{g(r)}~dr
\Big),~~~~~~~~~~~~~~~~~~~~~~~~~~~~~~
\end{equation}
\begin{equation}       \nonumber
\omega^{1} = - \frac{g^{\prime}(r)}{\zeta(r)}~dr, ~~~~~~~~~~~
\omega^{2} =g(r) (C_{2}dt + d\varphi) + f(r)~dr,~~~~~~~~~~~~~~~~~~~~~~~~~~~~~~~
\end{equation}
\begin{equation} \label{L47}
A^{0}(r) =\frac{k}{\Lambda} \Big(\zeta(r) C_{2} dt +\frac{ f(r)
\zeta(r)}{g(r)}~dr  + ( N(r) + \zeta(r) )~d\varphi
\Big),~~~~~~~~~~~~~~~~~~~~~~~~~~~~~~~
\end{equation}
\begin{equation}       \nonumber
A^{1}(r) =-\frac{k}{\Lambda}  \frac{g^{\prime}(r) }{
\zeta(r)}~dr, ~~~~~~~~ A^{2}(r) = \frac{k}{\Lambda} \Big(
(-\Lambda r + C_{2} ~ g(r))~dt  + f(r)~dr + g(r) ~d\varphi \Big),
\end{equation}
where
\begin{equation}       \nonumber
\zeta(r)=\sqrt{g^{2}(r)+ C_{1}},
\end{equation}
$C_{1}$ and $C_{2}$ are constants; and the $f(r)$ and $g(r)\neq0$ are arbitrary functions of ~$r$.
As in the previous section, varying this model gives the boundary term as follows:
\begin{equation} \nonumber
\delta I= -\frac{1}{4\pi}~ \delta \int_{\partial \mathcal{M}_{r}}
d^{2}x~  \epsilon^{ij} \Big[a~ \omega_{ic}\omega_{j}^{~c} + 2b~
e_{ic}\Big(\omega_{j}^{~c}-\frac{\Lambda}{k}A_{j}^{~c}\Big) + 2d~
\omega_{ic}A_{j}^{~c} -d~\frac{\Lambda}{k} A_{ic}A_{j}^{~c} +
\Lambda d~ e_{ic}e_{j}^{~c} \Big]
\end{equation}
\begin{equation} \label{L48}
+ \int_{\partial \mathcal{M}_{r}} d^{2}x~ \frac{\delta
I_{ct}}{\delta \gamma_{ij}}  \delta
\gamma_{ij},~~~~~~~~~~~~~~~~~~~~~~~~~~~~~~~~~~~~~~~~~~~~~~~~~~~~~~~~~~~~~~~~~~~~~~~~~~~~~~~~~~~
\end{equation}
such that for above solution of this model we have $I_{ct}=-\frac{b\sqrt{-\Lambda}}{2\pi}~  \sqrt{-\gamma}$.
Then, after some calculations we get the regularized quasilocal stress tensor for this model as:
\begin{equation} \label{L49}
T^{ij}= -\frac{b}{2\pi \sqrt{-\gamma}} \epsilon^{in}
\Big(\omega_{n}^{~c}  -\frac{\Lambda}{k} A_{n}^{~c} \Big)
\gamma^{jk} e_{kc} -\frac{b}{2\pi} \sqrt{-\Lambda}~ \gamma^{ij},
\end{equation}
where $\sqrt{-\gamma}=r N(r)$. Now, using above solution, we obtain
the components of quasilocal stress tensor as follows:
\begin{equation} \label{L50}
T^{00}= -\frac{b}{2\pi rN(r)}+\frac{b \sqrt{-\Lambda}}{2\pi
N^{2}(r)},~~~~~~~~~~~  T^{02}=T^{20}=0,~~~~~~~~~~ T^{22}=-\frac{b
\Lambda}{2\pi rN(r)}-\frac{b \sqrt{-\Lambda}}{2\pi r^{2}}.
\end{equation}

\subsubsection  {BTZ-type solution}
For the BTZ type ansatz \eqref{L27}, after some calculation we obtain the following solution:
\begin{equation}          \nonumber
N^2(r) = -M -\Lambda r^{2}+ \frac{J^{2}}{4 r^2} ,~~~~~~   N^{\phi}(r) =- \frac{J}{2 r^2},~~~~~~~~~~~~~~~~~~~~~~~
\end{equation}
\begin{equation}          \nonumber
\omega^{0}(r) =\xi(r) (D_{2} dt + d\varphi) + \rho(r)~dr,~~~~~~~~~~~~~~~~~~~~~~~~~~~~~~~~~~~~~~~
\end{equation}
\begin{equation}          \nonumber
\omega^{1}(r) = g(r)(D_{2} dt + d\varphi) + f(r)~dr,~~~~~~~~~~~~~~~~~~~~~~~~~~~~~~~~~~~~~~~
\end{equation}
\begin{equation}          \nonumber
\omega^{2}(r) = h(r)(D_{2} dt + d\varphi) + \sigma(r)~dr,~~~~~~~~~~~~~~~~~~~~~~~~~~~~~~~~~~~~~~~
\end{equation}
\begin{equation} \label{L51}
A^{0}(r) =\frac{k}{\Lambda} \Big( D_{2} \xi(r) ~dt
+ \rho(r)~dr +\Big( \xi(r) -N(r) ~\Big)~d\varphi \Big),~~~~~~~~~~~~~~
\end{equation}
\begin{equation}          \nonumber
A^{1}(r) =\frac{k}{\Lambda} \Big( \Big( \Lambda r +D_{2}~ g(r) \Big)~dt
+f(r)~dr + \Big( g(r) - rN^{\phi}(r) \Big)~d\varphi \Big),
\end{equation}
\begin{equation}          \nonumber
A^{2}(r) =\frac{k}{\Lambda} \Big( h(r)(D_{2} dt + d\varphi)  +
\Big( \sigma(r) +\frac{N^{\phi}(r)}{N(r)} \Big)~dr
\Big),~~~~~~~~~~~~~~~~~
\end{equation}
where
\begin{equation}          \nonumber
\xi(r)= \sqrt{g^{2}(r)+h^{2}(r)+D_{1}},~~~~~~~~~~~~~~~~~
\rho(r)= \frac{h^{\prime}(r) + f(r) \xi(r) }{g(r)},
\end{equation}
\begin{equation}          \nonumber
\sigma(r)= \frac{g(r) g^{\prime}(r) + h(r) h^{\prime}(r)}{g(r)
\xi(r) }  + \frac{f(r) ~
h(r)}{g(r)},~~~~~~~~~~~~~~~~~~~~~~~~~~~~~~~~
\end{equation}
$D_{1}$, $D_{2}$, $J$ and $M$ are constants; and the $f(r)$,
$g(r)$ and $h(r)$ are arbitrary functions of ~$r$. Similar to
\eqref{L49} the regularized quasilocal stress tensor for the
above   solution of this model has the following form:
\begin{equation} \label{L52}
T^{ij}= -\frac{b}{2\pi \sqrt{-\gamma}} \epsilon^{in}
\Big(\omega_{n}^{~c}  -\frac{\Lambda}{k}A_{n}^{~c} \Big)
\gamma^{jk} e_{kc} +\frac{b}{2\pi} \sqrt{-\Lambda} \gamma^{ij},
\end{equation}
where the counter term is $I_{ct}=\frac{b\sqrt{-\Lambda}}{2\pi}~  \sqrt{-\gamma}$.
Now, using the above solution we obtain the components of quasilocal stress tensor as follows:
\begin{equation} \nonumber
T^{00}= \frac{b}{2\pi rN(r)}-\frac{b \sqrt{-\Lambda}}{2\pi
N^{2}(r)},~~~~~~~~~~~~~~~~~~~~  T^{02}=T^{20}=\frac{b}{2\pi}
\sqrt{-\Lambda} \frac{N^{\phi}(r)}{N^{2}(r)},
\end{equation}
\begin{equation} \label{L53}
T^{22}=\frac{b \Lambda}{2\pi rN(r)}+\frac{b}{2\pi}
\sqrt{-\Lambda}
(\frac{N^{2}(r)-r^{2}(N^{\phi}(r))^{2}}{r^{2}N^{2}(r)}).~~~~~~~~~~~~~~~~~~~~~~~~~~~~~~~
\end{equation}
Then, the conserved charges, namely mass and angular momentum are determined as follows:
\begin{equation} \label{L54}
m=\frac{b}{2} M ,~~~~~~~ P_{\varphi}=\frac{b}{2}~ J.
\end{equation}
This metric has two singularities at
\begin{equation} \label{L55}
r_{\pm}= \sqrt{\frac{-M}{2 \Lambda} \Big( 1\mp\sqrt{1 +\frac{\Lambda J^{2}}{M^{2}}} ~\Big)}.
\end{equation}
where $r_{+}$ and $r_{-}$ are called event horizon and inner
horizon, respectively.  These singularities are the coordinate
singularities for which the Kretschmann scalar is
$K=R_{\mu\nu\rho\sigma} R^{\mu\nu\rho\sigma}=12 \Lambda^{2}$.
They describe a rotating $(J \neq 0)$ BTZ-like black hole in
$2+1$ dimensional spacetime with mass $M$ and angular momentum
$J$ which interact with non-abelian gauge fields $A_{i}^{~a}$.

\section  {$Ads/CFT$ correspondence for Chern-Simons action with semi-simple extension of Poincar\'{e} gauge group}
In this section, similar to \cite{J.D. Brown} and \cite{O.
coussaert}  we investigate the $Ads/CFT$ correspondence at the
boundary for the Chern-Simons action with semi-simple extension
of the Poincar\'{e} gauge group. Let us define the new generators
for the algebra of this group as follows:
\begin{eqnarray} \label{L56}
W_{a}^{\pm}=\frac{1}{2} \Big( -\frac{k}{\Lambda} Z_{a} \pm \frac{1}{\sqrt{-\Lambda}} P_{a} \Big),~~~~~~~~
\overline{W}_{a}=J_{a}+\frac{k}{\Lambda} Z_{a},
\end{eqnarray}
such that the commutation relations for this Lie algebra by use of \eqref{L35} have the following form:
\begin{eqnarray} \label{L57}
[W_{a}^{\pm},W_{b}^{\pm}] = \epsilon_{abc}W^{\pm c},~~~~~
[\overline{W}_{a},\overline{W}_{b}] = \epsilon_{abc}\overline{W}^{c},~~~~~
[W_{a}^{+},W_{b}^{-}] = 0,~~~~~
[W_{a}^{\pm},\overline{W}_{b}] = 0.
\end{eqnarray}
In this sence, we see that the semi-simple extension of the
Poincar\'{e} algebra  \textbf{$\mathcal{S}$}, is isomorphic to
the direct sum of three $so(2,1)$ Lie algebras i.e.
$\cal{S}\equiv$ $so(2,1) \oplus so(2,1) \oplus so(2,1)$.
Therefore, the gauge fields with these new generators have the
following forms:
\begin{eqnarray} \label{L58}
h_{i}=h_{i}^{+ a}~ W_{a}^{+} + h_{i}^{- a}~ W_{a}^{-} + \overline{h}_{i}^{~a}~ \overline{W}_{a},
\end{eqnarray}
where
\begin{eqnarray} \label{L59}
h_{i}^{\pm a}= \omega_{i}^{~a} \pm \sqrt{-\Lambda} e_{i}^{~a} -\frac{\Lambda}{k} A_{i}^{~a},~~~~~~~~~~~~~~
\overline{h}_{i}^{~a}= \omega_{i}^{~a}.
\end{eqnarray}

By choosing ~$x^{\pm}=\sqrt{-\Lambda}~t \pm \varphi =
\frac{t}{\ell}  \pm \varphi$~ and $C_{2}=\sqrt{-\Lambda}$, the
$Ads$ solution \eqref{L47} can be rewritten as:
\begin{equation}\label{L60}
h^{\pm}=\frac{1}{2}  \left( \begin{tabular}{cc}
                  $-\eta^{\pm}(r) dr$     &    $-y^{\pm}(r) dx^{\mp}$       \\
                  $-y^{\mp}(r) dx^{\mp}$   &   $\eta^{\pm}(r) dr$        \\
                \end{tabular} \right),~~~~~~~~~~~~~
\end{equation}

\begin{equation}\label{L61}
\overline{h}=  \frac{1}{2} \left( \begin{tabular}{cc}
                  $-\frac{g^{\prime}(r) }{\zeta(r)} dr$     &     $s^{+}(r) \Big( dx^{+}+\frac{ f(r)}{g(r)}dr\Big)$    \\
                  $s^{-}(r) \Big( dx^{+}+\frac{ f(r)}{g(r)}dr\Big)$      &    $\frac{g^{\prime}(r) }{\zeta(r)} dr$     \\
                  \end{tabular} \right),
\end{equation}
where
\begin{equation} \nonumber
\eta^{\pm}(r)= N^{-1}(r) +(-1 \pm \sqrt{-\Lambda}) \frac{g^{\prime}(r) }{\zeta(r)},
\end{equation}
\begin{equation} \nonumber
y^{\pm}(r)= \sqrt{-\Lambda}~ r \pm N(r),~~~~~~~~~~~~~~~~~
\end{equation}
\begin{equation}\nonumber
s^{\pm}(r)= g(r) \pm \zeta(r).~~~~~~~~~~~~~~~~~~~~~~~
\end{equation}
From \eqref{L60} and \eqref{L61} we see that ~$h_{+}^{+}=h_{-}^{-}=\overline{h}_{-}=0,$~
then we have
\begin{eqnarray}\label{L62}
h_{0}^{+}=-\sqrt{-\Lambda} ~ h_{2}^{+},~~~~~~
h_{0}^{-}=\sqrt{-\Lambda} ~  h_{2}^{-},~~~~~~
\overline{h}_{0}=\sqrt{-\Lambda} ~ \overline{h}_{2}.
\end{eqnarray}
In this case, the $2+1$ dimensional gravity model with semi-simple
extension of  Poincar\'{e} gauge Algebra \eqref{L40} can be
written as the sum of three Chern-Simons actions
\begin{eqnarray}\nonumber
I = K^{+} I_{cs}(h_{i}^{+}) + K^{-} I_{cs}(h_{i}^{-}) + \overline{K}~ I_{cs}(\overline{h}_{i}),~~~~~~~~~~~~
\end{eqnarray}
\begin{eqnarray}\label{L63}
K^{\pm}=\frac{1}{2}\Big( -\frac{k}{\Lambda} d \pm \frac{b}{\sqrt{-\Lambda}}  \Big),~~~~~~~~~~
\overline{K}=\Big( a +\frac{k}{\Lambda} d \Big),~~~
\end{eqnarray}
where $K^{\pm}$ and $\overline{K}$ are levels of the Chern-Simons
actions,  and actions $I_{cs}(h_{i}^{\pm})$ and
$I_{cs}(\overline{h}_{i})$ up to a surface term can be written as:
\begin{eqnarray}\label{L64}
I_{cs}(h_{i}^{\pm})=I^{\pm}=\frac{1}{4\pi} \int d^{3}x \Big[
h_{2}^{\pm a}  \partial_{0} h_{1a}^{\pm} - h_{1}^{\pm a}
\partial_{0} h_{2a}^{\pm} + 2 h_{0}^{\pm c} F_{12a}^{\pm} \Big],
\end{eqnarray}
\begin{eqnarray}\label{L65}
I_{cs}(\overline{h}_{i})=\overline{I}=\frac{1}{4\pi} \int d^{3}x
\Big[ \overline{h}_{2}^{~a} \partial_{0} \overline{h}_{1a} -
\overline{h}_{1}^{~a} \partial_{0} \overline{h}_{2a} + 2
\overline{h}_{0}^{~c} \overline{F}_{12a} \Big],~~~~~
\end{eqnarray}
such that the standard curvatures are
\begin{eqnarray}\label{L66}
F_{12a}^{\pm}= \partial_{1} h_{2a}^{\pm} - \partial_{2}
h_{1a}^{\pm}  + \epsilon_{abc} h_{1}^{\pm b} h_{2}^{\pm c} ,
\end{eqnarray}
\begin{eqnarray}\label{L67}
\overline{F}_{12a}= \partial_{1} \overline{h}_{2a} -
\partial_{2}  \overline{h}_{1a} + \epsilon_{abc}
\overline{h}_{1}^{b} \overline{h}_{2}^{c}.~~~
\end{eqnarray}
The variations of each of these Chern-Simons actions at the
boundary  $(r \rightarrow \infty)$ is not zero, even when
equations of motion hold, because of $\int d^{3}x ~Tr(h_{0}
\partial_{1} h_{2})$  term. This term along with conditions
$h_{+}^{+}=h_{-}^{-}=\overline{h}_{-}=0,$  at the boundary yields:
\begin{eqnarray}\label{L68}
\int d^{3}x ~ Tr(h_{0}^{A} \partial_{1} h_{2}^{A}) =
\frac{(-1)^{\delta_{+,A}}}{2}  \sqrt{-\Lambda} \int_{\Sigma} dt
d\varphi ~ Tr \Big[(h_{2}^{A})^2 \Big],
\end{eqnarray}
where $h_{i}^{A}=\{ h_{i}^{\pm a} W_{a}^{\pm},
\overline{h}_{i}^{~a} \overline{W}_{a} \}$  and $\Sigma$
demonstrates the two dimensional boundary. Then, the variations of
the model on the boundary is given by
\begin{eqnarray}\label{L69}
\delta \Big[\frac{1}{4\pi} \sqrt{-\Lambda} ~
Tr\Big(-K^{+}(h_{2}^{+})^2 + K^{-}(h_{2}^{-})^2  + \overline{K}
(\overline{h}_{2})^2\Big) \Big],
\end{eqnarray}
and in order to have $\delta I =0$, one must add this surface term with minus sign to the
action. Therefore, we have the following improved model:
\begin{eqnarray} \label{L70}
I = K^{+} I_{cs}(h_{i}^{+}) + K^{-} I_{cs}(h_{i}^{-}) +
\overline{K}~ I_{cs}(\overline{h}_{i})  -
\frac{1}{4\pi}\sqrt{-\Lambda} ~ \int_{\Sigma} dt d\varphi ~
Tr\Big[-K^{+}(h_{2}^{+})^2 + K^{-}(h_{2}^{-})^2 + \overline{K}
(\overline{h}_{2})^2\Big],
\end{eqnarray}
such that according to \eqref{L64} and \eqref{L65} the gauge
fields $h_{0}^{\pm}$ and  $\overline{h}_{0}$ have the role of
Lagrange multipliers, and variations of the model with respect to
these gauge fields yield the following constraints:
\begin{eqnarray}\label{L71}
F_{12}^{\pm} = \overline{F}_{12}= 0.
\end{eqnarray}
One solution for these constraints is ~ $h_{i}^{\pm} =
\overline{h}_{i} = 0$,~  then their gauge transformations $(h
\rightarrow g^{-1}dg+g^{-1}hg)$ are also a solution for the above
constraints, and we have
\begin{eqnarray}\label{L72}
h_{i}^{+} = G_{1}^{-1} \partial_{i} G_{1},~~~~~~   h_{i}^{-} =
G_{2}^{-1} \partial_{i} G_{2},~~~~~~       \overline{h}_{i}=
G_{3}^{-1} \partial_{i} G_{3},
\end{eqnarray}
where $G_{1}$, $G_{2}$ and $G_{3}$  must have the following
forms,  such that the radial components of the gauge fields
$h_{1}^{\pm}$ and $\overline{h}_{1}$ coincide with that of
\eqref{L60} and \eqref{L61} for the selection $f(r)=0$; i.e.
\begin{equation}\label{L73}
G_{1}(t,r,\varphi)= g_{1}(t,\varphi) \left( \begin{tabular}{cc}
                  $ U_{1}(r) $     &    $0$      \\
                  $0$     &   $ \frac{1}{U_{1}(r)} $       \\
                \end{tabular} \right),
\end{equation}
\begin{equation}\label{L74}
G_{2}(t,r,\varphi)= g_{2}(t,\varphi) \left( \begin{tabular}{cc}
                  $ U_{2}(r) $     &    $0$      \\
                  $0$     &   $ \frac{1}{U_{2}(r)} $       \\
                \end{tabular} \right),
\end{equation}
\begin{equation}\label{L75}
G_{3}(t,r,\varphi)= g_{3}(t,\varphi) \left( \begin{tabular}{cc}
                  $ U_{3}(r) $     &    $0$      \\
                  $0$     &   $ \frac{1}{U_{3}(r)} $       \\
                \end{tabular} \right),
\end{equation}
where $g_{1}(t,\varphi)$,~$g_{2}(t,\varphi)$~ and
$g_{3}(t,\varphi)$ are  arbitrary elements of the Lie group
$SO(2,1)$ and functions $U_{1}(r)$,~ $U_{2}(r)$ and $U_{3}(r)$
have the following forms
\begin{equation}\label{L76}
U_{1}(r)= \Big( y^{+}(r) \Big)^{\frac{-1}{\sqrt{-\Lambda}}}
~\Big( \sqrt{-\Lambda}~ s^{+}(r) \Big)^{(1-\sqrt{-\Lambda})},
\end{equation}
\begin{equation}\label{L77}
U_{2}(r)= \Big( y^{+}(r) \Big)^{\frac{-1}{\sqrt{-\Lambda}}}
~\Big( \sqrt{-\Lambda}~ s^{+}(r) \Big)^{(1+\sqrt{-\Lambda})},
\end{equation}
\begin{equation}\label{L78}
U_{3}(r)= \Big( \sqrt{-\Lambda}~ s^{+}(r) \Big)^{-1}.~~~~~~~~~~~~~~~~~~~~~~~~
\end{equation}
Using the above values for $G_{1}$, $G_{2}$ and $G_{3}$, one can write the surface term in \eqref{L70} as:
\begin{equation}\label{L79}
- \frac{1}{4\pi} \sqrt{-\Lambda} ~ \int_{\Sigma} dt d\varphi ~
Tr\Big[ -K^{+} \Big(g_{1}^{^{-1}}\partial_{2}~g_{1}\Big)^{2} +
K^{-} \Big(g_{2}^{^{-1}}\partial_{2}~g_{2}\Big)^{2} +
\overline{K} \Big(g_{3}^{^{-1}}\partial_{2}~g_{3}\Big)^{2} \Big],
\end{equation}
then the model \eqref{L70} can be rewritten as:
\begin{equation}\label{L80}
I= K^{+} S_{WZW}^{L}[g_{1}] + K^{-} S_{WZW}^{R}[g_{2}] + \overline{K}~ S_{WZW}^{R}[g_{3}],
\end{equation}
where $S_{WZW}^{L}[g_{1}]$, $S_{WZW}^{R}[g_{2}]$ and
$S_{WZW}^{R}[g_{3}]$ are  chiral WZW actions over $SO(2,1)$ such
that they describe a left-moving group element $g_{1}(x^{-})$ and
two right-moving group elements $g_{2}(x^{+})$ and $g_{3}(x^{+})$
respectively. Using the light cone coordinates
$\partial_{\pm}=\frac{1}{2}(\frac{1}{\sqrt{-\Lambda}}\partial_{0}\pm
\partial_{2})$ and~
$\partial_{+}g_{1}=\partial_{-}g_{2}=\partial_{-}g_{3}=0$, we have
\begin{equation}\label{L81}
S_{WZW}^{L}[g_{1}]= - \frac{1}{8\pi}  \int_{\Sigma} dt d\varphi
~  Tr\Big[ \dot{g}_{1}g'_{1} - \sqrt{-\Lambda} (g'_{1})^2 \Big] +
\Gamma[g_{1}],
\end{equation}
\begin{equation}\label{L82}
S_{WZW}^{R}[g_{2}]= - \frac{1}{8\pi}  \int_{\Sigma} dt d\varphi
~  Tr\Big[ \dot{g}_{2}g'_{2} + \sqrt{-\Lambda} (g'_{2})^2 \Big] +
\Gamma[g_{2}],
\end{equation}
\begin{equation}\label{L83}
S_{WZW}^{R}[g_{3}]= - \frac{1}{8\pi} ~ \int_{\Sigma} dt d\varphi
~  Tr\Big[ \dot{g}_{3}g'_{3} + \sqrt{-\Lambda} (g'_{3})^2 \Big] +
\Gamma[g_{3}],
\end{equation}
where $ \dot{g}_{i}=g_{i}^{-1}\partial_{0}~g_{i} $, $
g'_{i}=g_{i}^{-1} \partial_{2}~g_{i}, (i=1,2,3) $ ~and the
$\Gamma[g]$'s are the usual WZ term of the WZW action, which
using relations
$\partial_{0}h_{1}^{\pm}=\partial_{0}\overline{h}_{1}=0$ and
$F_{12}^{\pm}=\overline{F}_{12}=0$ can be written as:
\begin{equation}\label{L84}
\Gamma[g_{1}]= \frac{1}{4\pi} \int d^{3}x ~ Tr\Big[G_{1}^{-1}
\partial_{1}G_{1}.G_{1}^{-1}\partial_{0}G_{1}.G_{1}^{-1}\partial_{2}G_{1}
\Big],
\end{equation}
\begin{equation}\label{L85}
\Gamma[g_{2}]= \frac{1}{4\pi} \int d^{3}x ~ Tr\Big[G_{2}^{-1}
\partial_{1}G_{2}.G_{2}^{-1}\partial_{0}G_{2}.G_{2}^{-1}\partial_{2}G_{2}
\Big],
\end{equation}
\begin{equation}\label{L86}
\Gamma[g_{3}]= \frac{1}{4\pi} \int d^{3}x ~ Tr\Big[G_{3}^{-1}
\partial_{1}G_{3}.G_{3}^{-1}\partial_{0}G_{3}.G_{3}^{-1}\partial_{2}G_{3}
\Big].
\end{equation}

In this way, we prove that the $2+1$ dimensional gravity as
Chern-Simons action with gauge group $S$ is equivalent to sum of
three Chern-Simons actions with gauge group $SO(2,1)$ such that
the model at the boundary is a $CFT$ which is the sum of three
chiral $WZW$ models over the group $SO(2,1)$. Of course, these
results are also expected because there exist a decomposition of
the algebra $\cal{S}$ in terms of three  $so(2,1)$ algebras
\eqref{L57}.

\subsection {Central charge of the $CFT$ at boundary}
In order to calculate the central charge $c$ of the $CFT$ at the
boundary we use the following formula \cite{Ph. D. Francesco},
\cite{M. Henningson}:
\begin{equation}\label{L87}
Tr(T^{ij})=-\frac{c}{24\pi}\mathcal{R},
\end{equation}
where $T^{ij}$ and $\mathcal{R}$ are the regularized stress
energy tensor and scalar curvature of the boundary surface.  In
the previous section, we have calculated $T^{ij}$ for the
Ads-type solution \eqref{L49}. On the other hand, for calculating
$\mathcal{R}$ we use the extrinsic curvature $\theta_{ij}$ of the
boundary metric $\gamma_{ij}$
\begin{equation}\label{L88}
\theta_{ij}=-\frac{1}{2\sqrt{g_{rr}}} \partial_{r} \gamma_{ij}.
\end{equation}
Now, by use of the Fefferman-Graham expansion of boundary metric \cite{C. Fefferman}
\begin{equation}\label{L89}
\gamma_{ij}=r^{2} \gamma^{(0)}_{ij}+\gamma^{(2)}_{ij} + O(\frac{1}{r^{2}}), ~~~~~~~~~~~~
\gamma^{(0)}=diag(\Lambda,1),
\end{equation}
we have
\begin{equation}\label{L90}
\theta_{ij}=-r N(r) \gamma^{(0)}_{ij} + \cdot\cdot\cdot ~.
\end{equation}
Using the inverse of boundary metric \eqref{L89} in the following form
\begin{equation}\label{L91}
\gamma^{ij}=\frac{1}{r^{2}} (\gamma^{(0)})^{ij} -\frac{1}{r^{4}}
(\gamma^{(2)})^{ij} + \cdot\cdot\cdot ~,
\end{equation}
we obtain the trace of extrinsic curvature as:
\begin{equation}\label{L92}
\theta=\gamma^{ij}\theta_{ij}=-\frac{2 N(r)}{r} +
\frac{N(r)}{r^{3}}  \gamma^{(0)^{ij}} \gamma^{(2)}_{ij}  +
\cdot\cdot\cdot ~ .
\end{equation}
Then, using the following identity \cite{M. Hasanpour}
\begin{equation}\label{L93}
G_{ij} n^{i} n^{j}=-\frac{1}{2} (\mathcal{R}+\theta_{ij} \theta^{ij}-\theta^{2}),
\end{equation}
where $G_{ij}$ is the Einstein tensor, and $n^{i}$ is the unit
outward  pointing normal vector to the boundary $\partial M_{r}$,
for the geometry \eqref{L25} and \eqref{L47} and
$n^{i}=\frac{1}{\sqrt{g_{rr}}} \delta^{i,r}$ we have
\begin{equation}\label{L94}
G_{ij} n^{i} n^{j} = \frac{N^{2}(r)}{r^{2}} +\cdot \cdot \cdot ~,
\end{equation}
such that we obtain the scalar curvature of boundary at infinity $(r\rightarrow\infty)$ as follows:
\begin{equation}\label{L95}
\mathcal{R}=-\frac{2 \Lambda}{r^{2}} \gamma^{(0)^{ij}} \gamma^{(2)}_{ij} +\cdot\cdot\cdot~.
\end{equation}
Furthermore, for \eqref{L89} we have
\begin{equation}\label{L96}
\frac{1}{\sqrt{-\gamma}}=\frac{1}{\sqrt{-det(r^{2}\gamma^{(0)}_{ij})}
~\Big( 1+\frac{1}{r^{2}} \gamma^{(0)^{ij}}
\gamma^{(2)}_{ij}+\cdot\cdot\cdot
\Big)^{\frac{1}{2}}}=\frac{1}{\sqrt{-\Lambda}~r^{2}} ~\Big( 1
-\frac{1}{2r^{2}} \gamma^{(0)^{ij}}
\gamma^{(2)}_{ij}+\cdot\cdot\cdot \Big),
\end{equation}
and the non-zero components of the quasilocal stress tensor \eqref{L50} turn out to be
\begin{equation}\nonumber
T^{00}= -\frac{b}{2\pi} \frac{1}{\sqrt{-\Lambda}~r^{2}} ~\Big(
1-\frac{1}{2r^{2}}  \gamma^{(0)^{ij}} \gamma^{(2)}_{ij}
+\cdot\cdot\cdot \Big) -\frac{b
\sqrt{-\Lambda}}{2\pi}~\gamma^{00},
\end{equation}
\begin{equation}\label{L97}
T^{22}=-\frac{b \Lambda}{2\pi} \frac{1}{\sqrt{-\Lambda}~r^{2}}
~\Big( 1-\frac{1}{2r^{2}} \gamma^{(0)^{ij}} \gamma^{(2)}_{ij}
+\cdot\cdot\cdot \Big) -\frac{b
\sqrt{-\Lambda}}{2\pi}~\gamma^{22},
\end{equation}
such that at boundary $(r\rightarrow\infty)$ we have
\begin{equation}\nonumber
Tr(T^{ij})=\gamma_{ij}T^{ij}=\gamma_{00}T^{00}+\gamma_{22}T^{22}~~~~~~~~~~~~~~~~~~~~~~~~~~~~~~~~~~~~~~~~~~~~~~~~~~
\end{equation}
\begin{equation}\label{L98}
=-\frac{b}{2\pi} (\frac{2\Lambda r^{2}
+\cdot\cdot\cdot}{\sqrt{-\Lambda}~r^{2}}) ~\Big(
1-\frac{1}{2r^{2}} \gamma^{(0)^{ij}} \gamma^{(2)}_{ij}
+\cdot\cdot\cdot \Big) -\frac{b}{\pi} \sqrt{-\Lambda}~,
\end{equation}
where we have used ~$\gamma_{00}=\Lambda r^{2}+\cdot\cdot\cdot$~
and  ~$\gamma_{22}=r^{2}+\cdot\cdot\cdot$~. Finally, we obtain the
trace of the $Ads$ stress tensor as follows:
\begin{equation}\label{L99}
Tr(T^{ij})=-\frac{b}{2\pi}\frac{\sqrt{-\Lambda}}{r^{2}} \gamma^{(0)^{ij}} \gamma^{(2)}_{ij} +\cdot\cdot\cdot~,
\end{equation}
Now putting \eqref{L95} and \eqref{L99} in \eqref{L87} one can
obtain the central charge as
\begin{equation}\label{L100}
c=\frac{6b}{\sqrt{-\Lambda}}.
\end{equation}
On the other hand, the action \eqref{L40} can be rewritten as:
\begin{equation}\label{L101}
I=\frac{(K^{+}-K^{-})}{2}~(I^{+}-I^{-}) +\frac{(K^{+}+K^{-})}{2}~(I^{+}+I^{-}) +\overline{K}~\overline{I},
\end{equation}
where
\begin{equation}\nonumber
I^{+}-I^{-}= \frac{\sqrt{-\Lambda}}{2\pi} \int_{M} d^{3}x \:
\epsilon^{ijk}~ \Big\{ e_{ic}~\Big( \partial_{j}{} \omega_{k}^{\
c}-\partial_{k}{} \omega_{j}^{\ c} +  \epsilon^{abc}{}
\omega_{ja} \, \omega_{kb} \Big) - \frac{1}{3} \Lambda \:
\epsilon^{abc}~ e_{ic} \, e_{ja} \, e_{kb}
\end{equation}
\begin{equation}\label{L102}
-\frac{\Lambda}{k}~e_{ic} \Big(\partial_{j}~ A_{k}^{\
c}-\partial_{k}~  A_{j}^{\ c} + 2~ \epsilon^{abc}~ \omega_{ja}~
A_{kb} -\frac{\Lambda}{k} ~ \epsilon^{abc}{} A_{ja} \, A_{kb}
\Big) \Big\},
\end{equation}
is nothing but Hilbert-Einstein action coupled to the gauge fields. Hence, for it's coefficient we must have
\begin{equation}\label{L103}
\frac{(K^{+}-K^{-})}{2}=\frac{1}{8G\sqrt{-\Lambda}},
\end{equation}
then from \eqref{L63} we obtain
\begin{equation}\label{L104}
b=\frac{1}{4G},
\end{equation}
such that we find the central charge \eqref{L100} of the model as
\begin{equation}\label{L105}
c=\frac{3\ell}{2G},
\end{equation}
which is the central charge related to the Hilbert-Einstein
action from Chern-Simons theory with gauge group $SO(2,2)$
\cite{1 E. Witten}, \cite{2 E. Witten}. The reason for this
coincidence is that the energy stress tensors for the
~$(I^{+}+I^{-})$~ and ~$\overline{I}$~ parts in \eqref{L101} are
zero. Now, one may have a question that: what is the
contribution of the gauge fields in our model and in the
calculation of the central charge? The answer is that although
the energy stress tensor of the Chern-Simons model with gauge
group $SO(2,2)$, has the form ~$T^{ij}= -\frac{b}{2\pi
\sqrt{-\gamma}} \epsilon^{in} \omega_{n}^{~c} \gamma^{jk}
e_{kc}$~, and that of our model \eqref{L40} is ~$T^{ij}=
-\frac{b}{2\pi \sqrt{-\gamma}} \epsilon^{in} \Big(\omega_{n}^{~c}
-\frac{\Lambda}{k}A_{n}^{~c} \Big) \gamma^{jk} e_{kc}$,~ but their
values are the same in two models. Indeed, we have a shift
~$\omega^{a}_{\mu} \rightarrow \omega^{a}_{\mu}
-\frac{\Lambda}{k} A^{a}_{\mu}$~ in the spin connection as in
\cite{1 J.A. de Azcarraga}. Then, in one hand the geometries of
the boundaries of these two models (i.e. $\gamma_{ij}$) are the
same and on the other hand the values of the stress tensor
are also the same in two models, and consequently we have the same
central charges for these models. This motivates a question: Are
there two different $2+1$ dimensional gravity models such that they have
the same $CFT$ at their boundaries? Indeed, in the following we
show that the answer is positive and that these two $2+1$ dimensional gravities (i.e.
Chern-Simons models with the semi-simple extension of Poincar\'{e}
gauge group and SO(2,2) \cite{1 E. Witten}) are dual to each
other (of course, for special values of the constants $a,b$ and $d$ of the ad-invariant metric).

We note that for arbitrary values of the constants $a,b$ and $d$
of the ad-invariant metric, there is no general map to relate the
$SO(2,2)$ Chern-Simons model to the Chern-Simons action with
semi-simple extension of the Poincar\'{e} gauge group
\eqref{L40}. However, by selecting
~$d=\frac{\sqrt{-\Lambda}}{k}~b$ ~($K^{-}=0$ using \eqref{L63}),
the Chern-Simons model with gauge group $SO(2,2)$ having
the following form:
\begin{equation}\label{L106}
\tilde{I}=\frac{1}{8\pi} \int_{M} d^{3}x ~\epsilon^{ijk} \Big\{
2b^{\prime} ~e_{ic} (D_{j} \omega_{k}^{~c}  -\frac{1}{3} \lambda
\epsilon^{abc} e_{ja} ~ e_{kb}) +a^{\prime}~\omega_{ic}
(\partial_{j}~ \omega_{k}^{\ c}-\partial_{k}~ \omega_{j}^{\ c} +
\frac{2}{3}~ \epsilon^{abc} \omega_{ja} ~ \omega_{kb})
+a^{\prime}~ \lambda ~ e_{ic}~D_{j} e_{k}^{~c} ~  \Big\},
\end{equation}
is dual to our model \eqref{L40}; i.e. the following map
\begin{equation}  \nonumber
e_{i}^{~a} \rightarrow ~ \Xi~ (e_{i}^{~a} +
\frac{\sqrt{-\Lambda}}{k} A_{i}^{~a}),~~~~~~~~~~~~
\end{equation}
\begin{equation} \label{L107}
\omega_{i}^{~a} \rightarrow ~ \omega_{i}^{~a} +
\frac{\sqrt{-\Lambda}}{2}~(e_{i}^{~a} + \frac{\sqrt{-\Lambda}}{k}
A_{i}^{~a}),
\end{equation}
with
\begin{equation}  \nonumber
\lambda = \frac{-\Lambda}{4~\Xi^2},~~~~~~~~ b^{\prime}=b,~~~~~~~~
a^{\prime}=a,
\end{equation}
transforms this model \eqref{L106} to our model \eqref{L40}, where
$a^{\prime}$ and $b^{\prime}$ are arbitrary constants of the
SO(2,2) ad-invariant metric and
\begin{equation}  \nonumber
\Xi=1-\frac{a \sqrt{-\Lambda}}{2 b}.
\end{equation}
Indeed, this map is a canonical transformation and one can see
that the following canonical Poisson-brackets and the Hamiltonian
related to the $SO(2,2)$ Chern-Simons model
\begin{equation}\label{L108}
\{(\tilde{\Pi}_{e})_{i}^{~a}(x) ~,~ e_{j}^{~b}(y)\}=
\{(\tilde{\Pi}_{\omega})_{i}^{~a}(x) ~,~ \omega_{j}^{~b}(y)\}=
\epsilon_{ij} \eta^{ab} \delta^{2}(x-y),
\end{equation}
\begin{equation}\nonumber
\tilde{H}= \int d^{3}x\Big( (\tilde{\Pi}_{e})^{i}_{~a}~ \partial_{t}e_{i}^{~a} + (\tilde{\Pi}_{\omega})^{i}_{~a}~ \partial_{t}\omega_{i}^{~a} \Big) - \tilde{I} ~~~~~~~~~~~~~~~~~~~~~~~~~~~~~~~~~~~~~
\end{equation}
\begin{equation}\label{L109}
= -\frac{1}{8\pi}\int d^{3}x \epsilon^{ij} \Big(
8b^{\prime}\omega_{ia}
\partial_{t} e_{j}^{~a} +4a^{\prime} (\omega_{ia}
\partial_{t} \omega_{j}^{~a} + \lambda e_{ia}
\partial_{t} e_{j}^{~a})  \Big) - \tilde{I},
\end{equation}
where
\begin{equation} \nonumber
(\tilde{\Pi}_{e})_{i}^{~a}=
\frac{\partial \mathcal{\tilde{L}}}{\partial (\partial_{t}e_{~a}^{i})}= -\frac{1}{2\pi}~\epsilon_{i}^{~j} (b^{\prime} \omega_{j}^{~a}+ \lambda a^{\prime} e_{j}^{~a}),
\end{equation}
\begin{equation} \nonumber
(\tilde{\Pi}_{\omega})_{i}^{~a}=
\frac{\partial \mathcal{\tilde{L}}}{\partial (\partial_{t}\omega_{~a}^{i})}= -\frac{1}{2\pi}~\epsilon_{i}^{~j} (b^{\prime} e_{j}^{~a}+ a^{\prime} \omega_{j}^{~a}),~
\end{equation}
\begin{equation}\nonumber
(\tilde{\Pi}_{e})_{0}^{~a}= (\tilde{\Pi}_{\omega})_{0}^{~a}=0,~~~~~~~~~~~~~~~~~~~~~~~~~~~
\end{equation}
are transformed to the following Poisson-brackets and the Hamiltonian
related to our model \eqref{L40}
\begin{equation}\label{L110}
\{(\Pi_{e})_{i}^{~a}(x) ~,~ e_{j}^{~b}(y)\}=
\{(\Pi_{\omega})_{i}^{~a}(x) ~,~ \omega_{j}^{~b}(y)\}=
\{(\Pi_{A})_{i}^{~a}(x) ~,~ A_{j}^{~b}(y)\}= \epsilon_{ij}
\eta^{ab} \delta^{2}(x-y),
\end{equation}
\begin{equation}\label{L111}
H=-\frac{1}{8\pi}\int d^{3}x \epsilon^{ij} \Big[
8b(\omega_{ia}-\frac{\Lambda}{k}A_{ia})
\partial_{t} e_{j}^{~a} +4a \omega_{ia}
\partial_{t} \omega_{j}^{~a} +4d (k e_{ia}
\partial_{t} e_{j}^{~a} -\frac{\Lambda}{k} A_{ia}
\partial_{t} A_{j}^{~a} +2\omega_{ia}
\partial_{t} A_{j}^{~a})  \Big] - I,
\end{equation}
where $\epsilon^{12}=+1,$ the indices $i,j=1,2$ are the spatial
indices, and
\begin{equation}\nonumber
(\Pi_{e})_{i}^{~a}=
\frac{\partial \mathcal{L}}{\partial (\partial_{t}e_{~a}^{i})}= -\frac{1}{2\pi}~\epsilon_{i}^{~j} \Big( b(\omega_{j}^{~a}-\frac{\Lambda}{k} A_{j}^{~a})+kd~e_{j}^{~a} \Big),~~
\end{equation}
\begin{equation}\nonumber
(\Pi_{\omega})_{i}^{~a}=
\frac{\partial \mathcal{L}}{\partial (\partial_{t}\omega_{~a}^{i})}= -\frac{1}{2\pi}~\epsilon_{i}^{~j} \Big( b~e_{j}^{~a}+a \omega_{j}^{~a}+d A_{j}^{~a} \Big),~~~~~~~
\end{equation}
\begin{equation}\label{L112}
(\Pi_{A})_{i}^{~a}=
\frac{\partial \mathcal{L}}{\partial (\partial_{t}A_{~a}^{i})}= -\frac{1}{2\pi}~\epsilon_{i}^{~j} \Big( d(\omega_{j}^{~a}-\frac{\Lambda}{k} A_{j}^{~a})-\frac{\Lambda}{k} b~e_{j}^{~a} \Big),
\end{equation}
\begin{equation}\nonumber
(\Pi_{e})_{0}^{~a}= (\Pi_{\omega})_{0}^{~a}= (\Pi_{A})_{0}^{~a}=0,~~~~~~~~~~~~~~~~~~~~~~~~~~~~~~
\end{equation}
are the conjugate momentums corresponding to the gauge fields
$h_{i}^{~a}=(e_{i}^{~a},\omega_{i}^{~a},A_{i}^{~a}),$ which
according to \eqref{L107} are transformed as
\begin{equation}\nonumber
(\tilde{\Pi}_{e})_{i}^{~a} ~\rightarrow~ \frac{1}{2\Xi} \Big(
(\Pi_{e})_{i}^{~a} -\sqrt{-\Lambda}~(\Pi_{\omega})_{i}^{~a}
+\frac{k}{\sqrt{-\Lambda}}(\Pi_{A})_{i}^{~a} \Big)
\end{equation}
\begin{equation}\label{L113}
(\tilde{\Pi}_{\omega})_{i}^{~a} ~\rightarrow~
(\Pi_{\omega})_{i}^{~a}.~~~~~~~~~~~~~~~~~~~~~~~~~~~~~~~~~~~~~~~~~~~~~~~
\end{equation}
If we require that the maps \eqref{L107} relate the equations of
motion for the $SO(2,2)$ Chern-Simons model to the equations of
motion \eqref{L44}-\eqref{L46}, we must place another restriction
on the constants of the ad-invariant metric as
$a=\frac{b}{\sqrt{-\Lambda}}$ ~($\overline{K}=0$ using
\eqref{L63}). Now, these results mean that the two different $2+1$
dimensional gravities with $Ads$ background, are dual to each
other for the special values of the constants $a,b$ and $d$
($d=\frac{\sqrt{-\Lambda}}{k}~b$~ and
~$a=\frac{b}{\sqrt{-\Lambda}}$), and in this way they have the same
$CFT$ at the boundary. Furthermore, from the quantization of the
levels of the Chern-Simons model \cite{2 E. Witten}  we conclude
that the ~$(K^{\pm},\overline{K})$~
 must be integer numbers. Then, from
\eqref{L63} we have
\begin{equation}\label{L114}
d=-\frac{\Lambda}{k} (K^{+}+K^{-}),~~~~~~~~
b=\sqrt{-\Lambda} (K^{+}-K^{-}),~~~~~~
a=K^{+}+K^{-}+\overline{K},
\end{equation}
i.e. the constants ~$a,b$ and $d$~ of the ad-invariant metric of
$\cal{S}$ have discrete values.\footnote{Note that from non
degeneracy of the ad-invariant metric we have $d\neq0$.}

\section {\large {\bf Conclusions}}
We have presented the nine dimensional Maxwell and the semi-simple
extension of the Poincar\'{e} algebras for $2+1$ dimensional
spacetime and obtained $2+1$ dimensional gravity (with cosmological
term) coupled to gauge fields by gauge symmetric models,
equivalent to Chern-Simons models over the mentioned gauge
groups. Some $Ads$ and BTZ type solutions for the equations of
motion for these models have been obtained. For the Chern-Simons
model with semi-simple extension of the Poincar\'{e} gauge group
we have shown that at the boundary, this model is equivalent to
$CFT$ model i.e. a sum of three $SO(2,1)$ WZW chiral
model.\footnote{After finishing this work, we noticed that some new works
was put in arXiv about the Chern-Simons models with both Maxwell and
semi-simple extension of the Poincar\'{e} gauge groups in $2+1$
dimensions (see \cite{P. Salgado} and \cite{J. Diaz}).} Then, we
show that the central charge of the $CFT$ is the same as that of
$CFT$ at the boundary of $Ads$ spacetime related to the
Chern-Simons model with gauge group $SO(2,2)$. Furthermore, we
show that these two $2+1$ dimensional gravities are dual to each
other i.e. there is a canonical transformation which transforms
one model to the other one. The study of string theory in these $Ads$
and BTZ backgrounds is an open problem. Also, the study of
Maxwell and semi-simple extension of the Poincar\'{e} algebra in
$1+1$ dimensional spacetime and the related models are other open
problems. Some of these problems are under our investigation.

\vspace{0.5cm}
\textbf{Acknowledgments:} We would like to express our heartfelt
gratitude to F. Darabi and  M.M. Sheikh-Jabbari for carefully reading the manuscript and
his useful comments.
This research was supported by a research fund No. 217D4310 from
Azarbaijan Shahid Madani university.


\end{document}